\documentclass[aps,prb,twocolumn,citeautoscript,superscriptaddress]{revtex4}
\usepackage{amsmath}
\usepackage{graphicx}
\usepackage{dcolumn}
\usepackage{float}
\setcitestyle{square}

\begin{document}

\title{Superconducting gap symmetry of the noncentrosymmetric superconductor W$_3$Al$_2$C}
\author{R. Gupta}
 \email{ritu.gupta@psi.ch}
\affiliation{Laboratory for Muon Spin Spectroscopy, Paul Scherrer Institute, CH-5232 Villigen PSI, Switzerland}
\author{T. P. Ying}
\affiliation{Materials Research Centre for Element Strategy, Tokyo Institute of Technology, Yokohama 226-8503, Japan}
\author{Y. P. Qi}
\affiliation{School of Physical Science and Technology, ShanghaiTech University, Shanghai 201210, China}
\author{H. Hosono}
\affiliation{Materials Research Centre for Element Strategy, Tokyo Institute of Technology, Yokohama 226-8503, Japan}
\author{R. Khasanov}
 \email{rustem.khasanov@psi.ch}
\affiliation{Laboratory for Muon Spin Spectroscopy, Paul Scherrer Institute, CH-5232 Villigen PSI, Switzerland}
\date{\today}

\begin{abstract}
A detailed zero-field and transverse-field muon spin relaxation/rotation ($\mu$SR) experiemnts have been carried out on the recently discovered non-centrosymmetric superconductor W$_3$Al$_2$C to speculate about its superconducting ground state. Bulk nature of superconductivity below 7.6 K is confirmed through magnetization measurements. No change in the $\mu$SR spectra collected above and below $T_c$ is visible, ruling out the possibility of spontaneous magnetic field below $T_c$. This confirms that time-reversal symmetry is preserved for W$_3$Al$_2$C upon entering in the superconducting ground state. Temperature dependent superfluid density [$\rho_s(T)$], which directly reflects the superconducting gap symmetry is obtained by the analysis of spectra obtained from the transverse-field $\mu$SR experiments. Despite a non-centrosymmetric structure, W$_3$Al$_2$C adopts a fully gaped spin-singlet superconducting ground state with a zero temperature value of gap $\Delta_0$ = 1.158(8) meV with gap-to-$T_c$ ratio 2$\Delta_0/k_BT_c\approx$3.54, classifying this material as a weakly-coupled superconductors. 

\end{abstract}

\maketitle

\section{INTRODUCTION}
In the quest of achieving superconductivity (SC) at high temperature, there has been enormous research on cuprates, Fe-based pnictide superconductors, heavy fermion superconductors etc \cite{Bednorz,Steglich,Kamihara}. These class of materials posses unconventional nature of SC which transcends from the expectations of the standard BCS model. The pairing of Cooper pairs in these superconductors is mediated by the charge/magnetic/valence fluctuations \cite{Mathur,Chen0,Morosan,Yuan0}, rather than the phonons as for the case of BCS superconductors. An important role is played by the crystal structure in deciding the pairing symmetry in unconventional superconductors. Most of the superconductors discovered so far posses a center of inversion in their crystal structure. The SC in these materials can be classified either as spin-singlet or spin-triplet type. No intermixing is permitted for such superconductors\cite{Anderson}. However, recent interests has been developed in a so-called non-centrosymmetric (NC) class of materials as they posses various exotic properties including unconventional SC, time reversal symmetry breaking (TRSB) in the superconducting state, topological protected surface states etc. Due to the lack of inversion symmetry, they generate an asymmetric spin-orbit coupling (ASOC) which lifts the degeneracy of the conduction band electrons and hence resulting in the splitting of Fermi surface, $i.e.$, splitting of spin-up and spin-down bands. As a result both inter- or intra- band Cooper pairs can be formed and hence admixture of spin-singlet and spin-triplet Cooper pairs are permitted.  The mix parity in case of NC superconductors can host complex superconducting properties.  

The research in the field of NC superconductors was triggered after the discovery of unusual superconducting ground state namely line nodes in heavy fermion compound CePt$_3$Si\cite{Bonalde}. Few other examples in this category are: CeIrSi$_3$\cite{Mukuda}, Mo$_3$Al$_2$C\cite{Bauer1,Bauer2}, Li$_2$Pt$_3$B\cite{Yuan,Nishiyama} exhibits line nodes, whereas few others such as LaNiC$_2$\cite{Chen}, (La,Y)$_2$C$_3$\cite{Kuroiwa}, LaPt$_2$Si$_2$ \cite{Das}etc. show multiband SC. The effect of ASOC has been directly observed in well acclaimed weakly correlated system Li$_2$(Pd,Pt)$_3$B. The parent compound  Li$_2$Pd$_3$B is a conventional BCS superconductor\cite{Rustem1} which gradually transforms from spin singlet to spin triplet SC by inclusion of Pt in place of Pd \cite{Yuan, Nishiyama, Badica}. The reason is attributed to the increase in strength of ASOC which is proportional to $Z^4$. The few other superconductors with strong ASOC where non-trivial superconducting ground state has been observed are: La$_7$Ir$_3$ \cite{Barker}, Re-based superconductors Re$_6$X (X = Zr, Hf, Ti)\cite{Singh, Singh1} etc. This implies that strength of ASOC plays an important role in deciding the superconducting gap symmetry. Very recently, NC Mo$_3$Al$_2$C superconductor ($\beta$ Mn-type structure, space group $P4_132$) has gained a significant attention due to its similar geometrical configuration to well established unconventional superconductor Li$_2$(Pd,Pt)$_3$B. Strong signatures of the SC deviating from standard-BCS behavior has been speculated via following observations: absence of Hebbel-Slichter peak, a power law behavior of spin-lattice relaxation rate measured through $^{27}$Al NMR, electronic specific heat\cite{Bauer1}.  Nodal type gap structure is suggested from the pressure enhanced $T_c$\cite{Bauer1}. In contrast, the microscopic techniques like $\mu$SR, Tunnel Diode Oscillator evidence a nodeless state of the superconducting order parameter\cite{Karki, Bonalde1, Bauer2, Hafliger}. It was proposed that nodal behavior was not observed in these techniques probably due to trace fraction of triplet SC. It is natural to expect that inducing stromger ASOC might increase the chances of unconventional SC with triplet pairing. 

To look for such a possibility, recently, Ying $et$ $al.$ were able to successfully grow single phase superconductor W$_3$Al$_2$C with $T_c\approx$7.5 K, where Mo was replaced by heavier element W to enhance ASOC \cite{Ying}. W$_3$Al$_2$C is isostructural to Mo$_3$Al$_2$C. Indeed, the first principal studies point towards the pronounced effect of ASOC on band structure and Fermi surface topology. Additionally, electronic specific heat was fitted with a power law expression, hinting towards a complex gap structure. However, the previous studies has been limited down to 2 K. There is a clear need of comprehensive microscopic techniques to explore the superconducting gap symmetry of W$_3$Al$_2$C. This motivated us to conduct muon spin relaxation/rotation ($\mu$SR) measurements to estimate temperature-dependent magnetic penetration depth [$\lambda(T)$], which in turn is proportional to the superfluid density  [$\lambda(T)\propto n_s^{-2}(T)$] hence directly reflecting the superconducting gap symmetry.   

This paper is organized in the following manner: Section II describes the sample-preparation and initial characterization procedure including the results of magnetization measurements, as well as the details about the $\mu$SR experiments. Section III is dedicated to results and discussion part involving TF and ZF $\mu$SR experiments. The conclusions are reported in Section IV. 

\section{Experimental details}
\subsection{Sample preparation and characterization}
The polycrystalline W$_3$Al$_2$C sample was synthesized by the high-pressure method with the detailed description being reported in the Ref. \cite{Ying} The optimal mole ratio of the elements W:Al:C = 3:1.8:0.8 were first ball milled for two days in a glove box to ensure homogeneity of the final product. The mixture was then pressed into pellets and sealed inside h-BN capsule. The pellets were then heated at 2173 K in an environment of  high pressure of 5 GPa for 24 h, followed by a slow cooling down for one day. The sample investigated in the present study with $\mu$SR is from the same batch, which has been used previously by Ying $et$ $al.$\cite{Ying}

To pre-characterize the superconducting properties of W$_3$Al$_2$C, temperature-dependent magnetic susceptibility $M(T)$ was measured in the field-cooled (FC) and zero-field-cooled (ZFC) mode in an applied field of 2 mT.  The magnetization experiments were performed using a commercial physical properties measuring system (PPMS, QUANTUM design magnetometer). It can be seen from Fig. 1, $M(T)$ data displays a sharp transition with width $\approx$ 0.1 K with the onset of superconductivity at $T_c$ =\text{ 7.5 K}. Bifurcation of FC and ZFC signal is an immanent feature of type-II superconductor with moderate to strong pinning, where magnetic flux is pinned upon cooling the system in an applied magnetic field. 
\begin{figure}[!htb]
\centering
\includegraphics[width=9 cm, keepaspectratio]{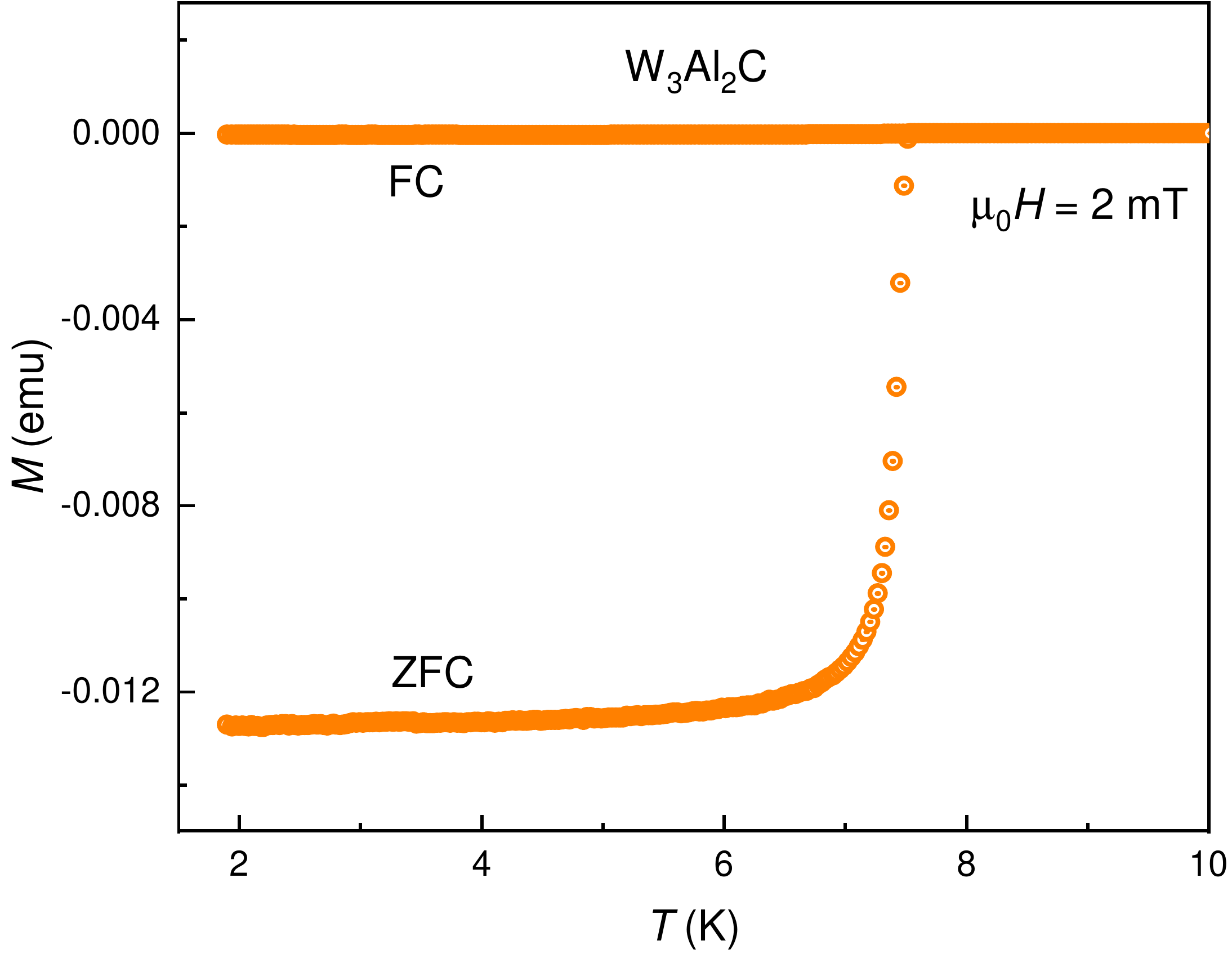}
\caption{\label{fig:rho@c} a) The temperature-dependent magnetization $M(T)$ of W$_3$Al$_2$C collected in an applied field of 2 mT with zero-field-cooled (ZFC) and field-cooled (FC) mode.}
\end{figure}

Zero field (ZF) and transverse field (TF) muon spin rotation/relaxation ($\mu$SR) experiments were carried out on GPS (General Purpose Surface) spectrometer, situated at the $\pi$E1 beamline, at the Paul Scherrer Institute (PSI), Villigen, Switzerland. The experiments were carried out in the temperature range 1.5 to 10 K in presence of desired magnetic field for TF and in zero field for the ZF configuration. For TF experiments, the sample was first cooled down from a temperature well above $T_c$  to the base temperature in presence of a magnetic field and then spectra were collected during warming up.  To minimize the statistical error bars, typical counting statics was kept at $\approx$10$^7$ positrons events for each data point. The TF and ZF $\mu$SR spectra were analyzed with the help of the free software MUSRFIT \cite{Suter_MuSRFit_2012}.

\section{results and discussion}
\subsection{TF $\mu$SR experiments}
\begin{figure*}[!htb]
\centering
\includegraphics[width=19 cm, keepaspectratio]{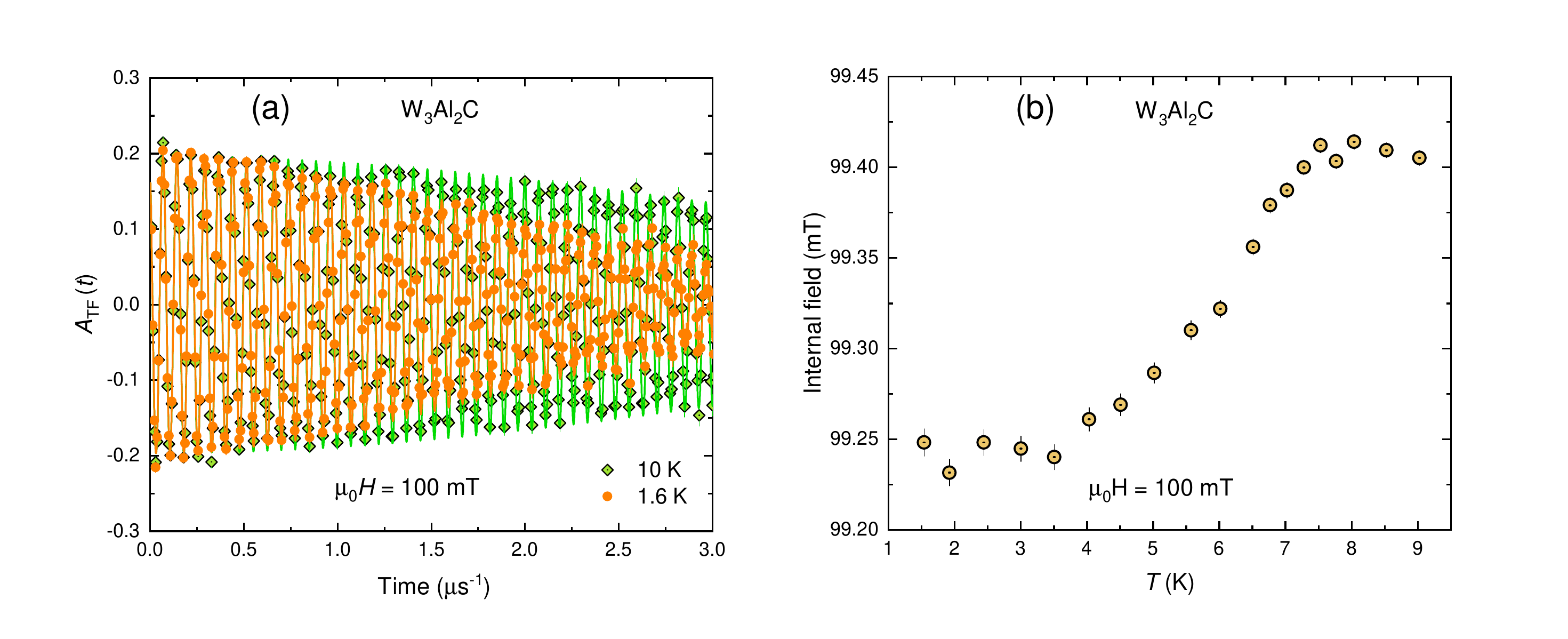}
\caption{\label{fig:rho@c} a) The TF-$\mu$SR time spectra of polycrystalline W$_3$Al$_2$C collected above $T_c$ (10 K, green symbols) and below $T_c$ (1.6 K, orange symbols) in a magnetic field of 100 mT. Solid lines through the data represent fitting using Eq. 1.  b) Internal magnetic field as a function of temperature in an applied field of 100 mT.}
\end{figure*}
To elucidate the nature of superconducting order parameter, we have carried out transverse field (TF) $\mu$SR experiments in the mixed superconducting state of W$_3$Al$_2$C with various applied magnetic fields: 30, 100, 300, and 600 mT. The TF-$\mu$SR asymmetry spectra were collected at several temperatures between 1.5 K to 10 K for GPS spectrometer. Fig. 2(a) displays two representative TF-$\mu$SR spectra in the normal state (10 K) and in the superconducting state (1.6 K), collected in a magnetic field of 100 mT. The TF-$\mu$SR signal in the superconducting state shows much faster damping compared to normal state due to inhomogeneous magnetic field distribution as a result of vortex lattice formation. The small damping above $T_c$ is due to the static nuclear magnetic moments. The $\mu$SR spectra collected was fitted to the following oscillatory decaying Gaussian function:
\begin{equation}
A\textsuperscript{TF}(t)=A\textsuperscript{TF} (0) \exp(-\sigma\textsubscript{tot}^2/2)\cos(\gamma_{\mu}B\textsubscript{int}t+\phi).
\end{equation}
Here, $A\textsuperscript{TF} (0)$ is the initial asymmetry belonging to the sample. $\gamma_{\mu}/2\pi$ = 135.5 MHz/T is the muon gyromagnetic ratio. $B\textsubscript{int}$ is the local internal magnetic field sensed by muons implanted in the sample, $\phi$ is the initial phase offset of the initital muon spin polarization with respect to positron detector. $\sigma\textsubscript{total}$ is the muon depolarization rate, which is comprised of the following two terms:  $\sigma\textsubscript{total}^2$ = $\sigma\textsubscript{sc}^2$+$\sigma\textsubscript{nm}^2$. Solid lines in the Fig. 2 (a) correspond to the fitting of observed spectra using Eq. 1. $\sigma\textsubscript{sc}$ and $\sigma\textsubscript{nm}$ are the muon depolarization rates associated with the flux-line lattice and the nuclear magnetic  moments, respectively.  $\sigma\textsubscript{nm}$ is expected to be intact in the entire temperature range and can be estimated by analyzing one of the spectra above $T_c$. Thus, we can extract $\sigma\textsubscript{sc}$ by quadratically subtracting $\sigma\textsubscript{nm}$ from $\sigma\textsubscript{total}$. The muon depolarization rate $\sigma\textsubscript{sc}$ is related to the penetration depth and hence to the superfluid density ($\sigma\textsubscript{sc}\propto\lambda^{-2}\propto n_s$). Thus, the superconducting gap symmetry can be deduced from the temperature dependence of $\sigma\textsubscript{sc}(T)$. Fig. 2(b) shows the temperature dependence of internal magnetic field sensed by the muon in presence of 100 mT magnetic field. The flux expulsion is clearly visible from the reduced value of internal magnetic field in the superconducting state.

\begin{figure}[!htb]
\centering
\includegraphics[width=9 cm, keepaspectratio]{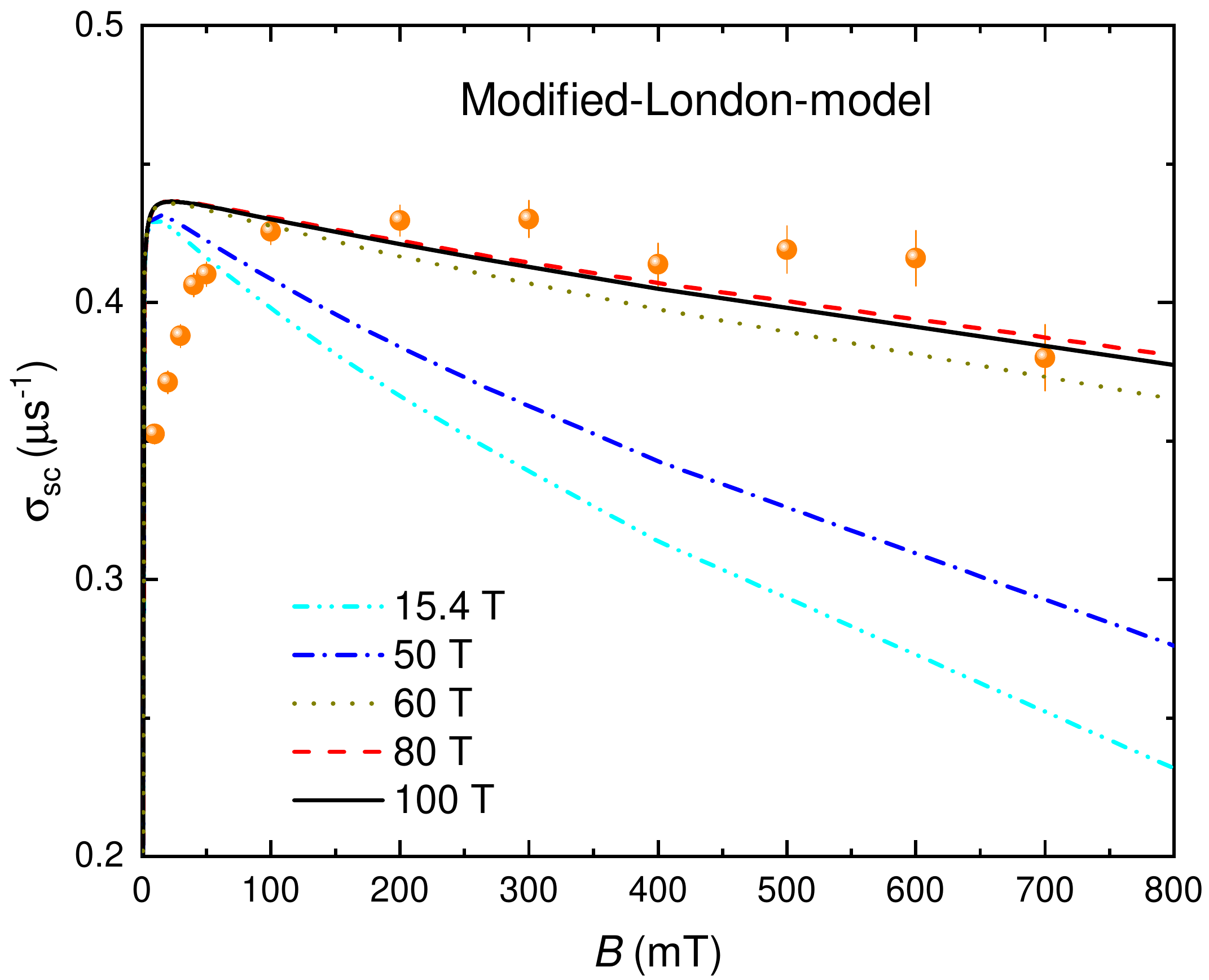}
\caption{\label{fig:rho@c}  The field dependence of muon depolarization rate $\sigma(B)$ analyzed with modified London model , Eq. 2, which has been adapted for single-gap $s$-wave superconductor. Various analyses were carried out by fixing upper critical field $\mu_0H_{c2}$ to 15.4, 50, 60, 80, and 100 T. Refer to the text for detailed description.}
\end{figure} 
\begin{figure*}[!htb]
\centering
\includegraphics[width=19 cm, keepaspectratio]{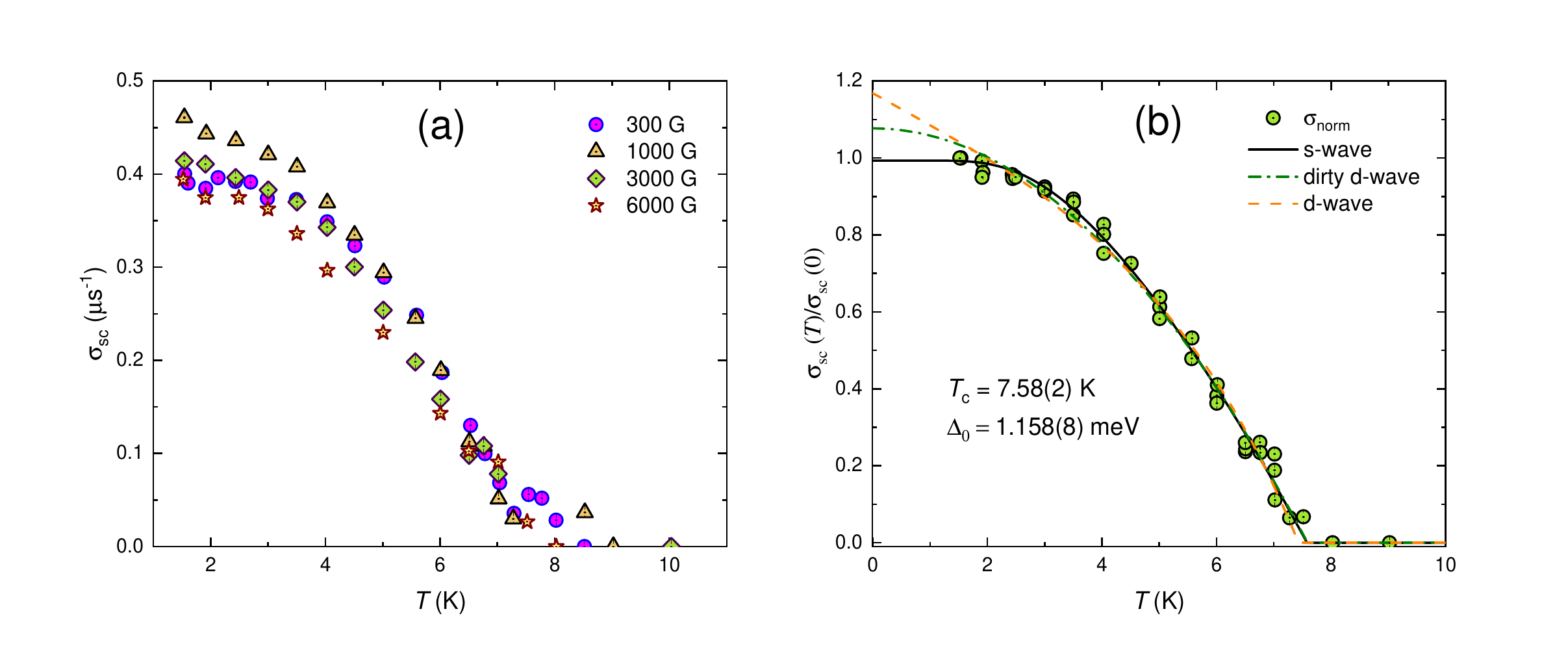}
\caption{\label{fig:rho@c} The temperature evolution of Gaussian depolarization rate $\sigma_{sc}(T)$ (symbols) a) for all four measured fields and b) combined for all fields with fits using a single gap $s$-wave model (solid black lines), a $d$-wave model (orange dotted line), and a dirty $d$-wave model approximated through a power law expression [1-$(T/T_c)^2$]. See text for details.}
\end{figure*}
The magnetic field and temperature evolution of $\sigma\textsubscript{sc}(T,B)$ has been obtained by fitting Eq. 1 to the experimentally observed asymmetry-time spectra.  Firstly, we have measured TF-spectra at base temperature 1.5 K in presence of various fields ranging from 0.01 T to 0.7 T. Fig. 3 represents the variation of muon depolarization rate as a function of field $\sigma_{sc}(B)$ at 1.5 K. The  $\sigma_{sc}(B)$ data was analyzed with the model presented by Serventi  $et$ $al.$ \cite{Serventi} In the model proposed, the second moment of magnetic field distribution within the flux-line-lattice (FLL) is calculated within the framework of modified London model with following expression:
\begin{equation}
\overline{\Delta B^2} =\bigg (\frac{\sigma\textsubscript{sc}^2}{\gamma_{\mu}}\bigg)^2 = B^2\sum_{q\neq 0}\bigg[\frac{e^{-q^2\xi^2/2(1-b)}}{1+q^2\lambda^2/(1-b)}\bigg]^2.
\end{equation}  
Here, $b=B/B_{c2}$ is the reduced magnetic field, with $B_{c2}$ as the upper critical field, $\xi$ is the coherence length, and $\lambda$ is the penetration depth.  $B$ is the mean field within FLL. The reciprocal lattice vector $q = 4\pi/\sqrt{a}(m\sqrt{3}/2, n+m/2)$, adapted for hexagonal flux line lattice, with $a$ being the intervortex distance, $m$ and $n$ are the integer numbers. The applied magnetic field $B_{app}\ll B_{c2}$ (predicted value of $B_{c2}$ = 15.4 T from previous study\cite{Ying}), hence it is difficult to speculate about the exact value of upper critical field from $\sigma(B)$ data. Moreover, we carried out the analysis of experimental $\sigma(B)$ data using Eq. 2 with five different values of the upper critical field: 15.4, 50, 60, 80, and 100 T. It can be seen from Fig. 3, $\sigma(B)$ data has very poor consistency with $B_{c2}$ = 15.4 T. The consistency becomes better as we go to higher $B_{c2}$ values. The theoretical curve with $B_{c2}$ = 50 T does not describe the data. However, as soon we approaches 60 T, consistency becomes better, with improving further for 80 and 100 T.  Although, we can not determine the absolute value of $B_{c2}$, but at least a lower limit on the upper critical field value, $i.e.$, 50 T $< B_{c2}$ can be described.

The temperature dependent $\sigma\textsubscript{sc}(T)$ for four different applied fields is shown in Fig. 4(a). Fig. 4 (b) shows the normalized muon depolarization rate $\sigma\textsubscript{sc}(T)/\sigma\textsubscript{sc}(0)$ as a function of temperature combined for the three fields 100, 300, and 600 mT, which are lying above the maximum in $\sigma (B)$. It can be seen that $\sigma\textsubscript{sc}(T)$ [Fig. 4(b)] tends to saturate below $\simeq T_c$/3. This rule out already the possibility of nodes in the superconducting energy gaps at the Fermi surface. However, to further speculate about the pairing symmetry quantitatively, we have employed different superconducting gap models to analyze the muon depolarization rate $\sigma\textsubscript{sc}(T)$. $\sigma\textsubscript{sc}$ can be calculated from the superconducting gap $\Delta (T,\phi)$ using standard local London approximation ($\lambda\ll \xi$)\cite{Tinkham}:
\begin{equation}
\frac{\sigma\textsubscript{sc}(T)}{\sigma\textsubscript{sc}(0)}=\frac{\lambda_L^{-2}(T)}{\lambda_L^{-2}(0)}=1+\frac{1}{\pi}\int_{0}^{2\pi}\int_{\Delta (T, \phi)}^{\infty}\frac{\partial f}{\partial E } \frac{EdEd\phi}{\sqrt{E^2-\Delta (T, \phi)^2}},
\end{equation}
with $f=[1+\text{exp}(E/k_BT)]^{-1}$ being the Fermi distribution function and $\phi$ is the azimuthal angle along the Fermi surface. $\Delta(T,\phi)$ = $\Delta(T)g_{\phi}$, where $\Delta(T)$ is the temperature dependence and $g_{\phi}$ is the angle dependence of gap function, with latter has a value $g_{\phi}$ = 1 for $s$-wave, and cos2$\phi$ for $d$-wave pairing symmetry.  The temperature dependence of the gap function $\Delta(T)$ is approximated in the standard way:  $\Delta(T)=\Delta_0\tanh\{1.821[1.018(T_c/T-1)^{0.51}]\}$\cite{Rustem_AuBe}, with $\Delta_0$ being the gap value at 0 K. Three different gap models namely $s$-wave without nodes, $d$-wave with line nodes, and a dirty $d$-wave model were considered for the analysis. Power law expression $1-(T/T_c)^2$ was tested, which has been proven theoretically to best describe the case of a dirty $d$-wave superconductor \cite{Hirschfeld}. Fig. 4(b) represents the temperature-dependent normalized muon depolarization rate fitted with the aforementioned gap models. It can be seen that momentum independent s-wave model is most compatible with the experimental data.  A poor agreement can be seen between the experimental data and $d$-wave models at low temperature, which rules out the plausibility of line nodes in the gap around the Fermi surface. The best described parameter values for s-wave model are: $\Delta_0$ = 1.158(8) meV and $T_c$ = 7.58(2) K. The superconducting transition temperature $T_c$ is similar to that obtained from the magnetization measurement. The superconducting gap to $T_c$ ratio $2\Delta_0/k_BT_c$ = 3.54, which is pretty close to the the universal BCS value 3.53, keeping this material in the list of weakly coupled superconductor. It should also be noted that previous specific heat analysis suggests the gap-to-$T_c$ ratio 5.43, much higher than obtained in this study. The possible reason of the difference in the two vlaues could be as the calculated density of states were used to estimate gap-to-$T_c$ ratio by Ying $et$ $al.$ \cite{Ying}, which might lead to slight error in the value.

\subsection{ZF $\mu$SR experiments}
\begin{figure}[!htb]
\centering
\includegraphics[width=9 cm, keepaspectratio]{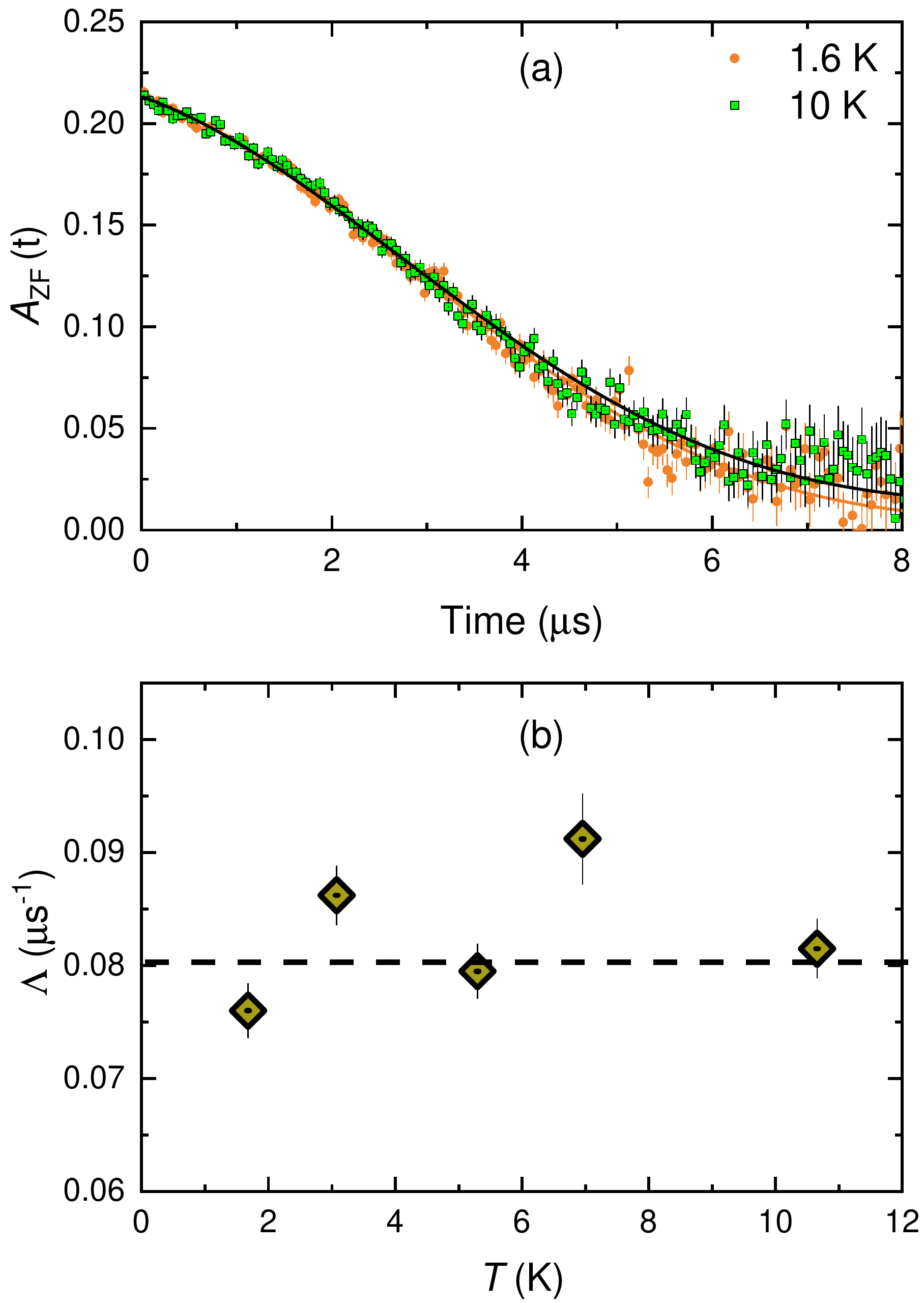}
\caption{\label{fig:rho@c} a) The representative ZF asymmetry-time spectra of polycrystalline W$_3$Al$_2$C measured at 1.6 K (orange symbol, well below $T_c$) and 10 K (green symbol, well above $T_c$). b) Temperature variation of electronic relaxation rate $\Lambda (T)$  obtained by fitting Eq. (4) to the corresponding datasets. The relaxation rate did not show distinguishable changes (within the error bars) across the superconducting transition $T_c$ = 7.6 K. The horizontal dashed line indicates the absence of change in the relaxation rate at $T_c$.}
\end{figure}
Muon spin rotation/relaxation ($\mu$SR) is a very sensitive technique which can probe extremely small magnetic field of the order of 10$^{-4}$ T. Consequently, in absence of any external applied field, even a tiny amount of spontaneous magnetic field which may arise due to time reversal symmetry breaking in the superconducting state can be detected. 

In order to search for possible magnetic field (static or fluctuating) in the superconducting state of W$_3$Al$_2$C, we collected various ZF-spectra in the temperature range from 1.6 K to 10 K. For representative manner, we have shown in Fig. 5(a), the ZF $\mu$SR spectra collected well below (1.6 K) and well above $T_c$ (10 K). The time dependent asymmetry [$A(t)$] collected above and below $T_c$ does not show distinct difference. No additional relaxation of the $\mu$SR signal in the superconducting state implies the absence of spontaneous internal field upon entering in the superconducting state. In fact, this observation suggests that time reversal symmetry is preserved despite of the non-centrosymmetric structure and stronger ASOC of W$_3$Al$_2$C. The observed asymmetry-time spectra is well described by a damped static Gaussian Kubo-Toyabe function:
\begin{equation}
A(t)=A_s\exp(-\Lambda t)G\textsubscript{KT}+A_\textsubscript{bg},
\end{equation}
where $G\textsubscript{KT}$ is the Gaussian Kubo-Toyabe function\cite{Kubo}, accounting for the the muon depolarization rate associated with the static randomly oriented local field due to nuclear magnetic moments. $A_s$ and $A_\textsubscript{bg}$ are the initial asymmetries associated to the sample and background, respectively. $\Lambda$ represents the electronic spin relaxation rate, which is additionally introduced in order to to account for any possible distribution of fields due to electronic spins. Gaussian Kubo Toyabe function $G\textsubscript{KT}$ has following functional form: 
\begin{equation}
G\textsubscript{KT}(t)=\frac{1}{3}+\frac{2}{3}(1-\sigma\textsubscript{ZF}^2t^2)\exp\bigg(-\frac{\sigma\textsubscript{ZF}^2t^2}{2}\bigg),
\end{equation}
where $\sigma\textsubscript{ZF}$ is the width of the nuclear dipolar field (the local field distribution $<B_{\mu}>$ = $\sigma/\gamma_{\mu}$, where muon gyromagnetic ratio $\gamma_{\mu}$ = 135.53 MHz/T) experienced by the muon-spin ensemble.  The Gaussian depolarization rate $\sigma\textsubscript{ZF}$ was fixed to a value 0.1893 $\mu$s$^{-1}$, estimated from fitting a spectra well above $T_c$. Fig. 5(b) displays the  temperature-dependent electronic relaxation rate $\Lambda(T)$. Within the experimental accuracy, no visible change in the relaxation was observed across $T_c$, excluding the plausibility of spontaneous internal magnetic fields which might break TRS in the superconducting state in this system. Therefore, we can safely conclude that time reversal symmetry is preserved in the NC W$_3$Al$_2$C upon entering in the superconducting ground state. Beside W$_3$Al$_2$C, there are few other examples of NC superconductors which did not show TRS breaking in the superconducting state $e.g.$ LaPt$_2$Si$_2$ \cite{Das}, BaPt$_3$Si \cite{Bauer3}, LaPt$_3$Si \cite{Smidman}, and many others. Thus, we can argue that TRS breaking is not an immanent feature of NC superconductors. 

\section{Conclusion}
The superconducting properties of non-centrosymmetric superconductor W$_3$Al$_2$C has been examined by means of magnetization, zero field and transverse field $\mu$SR experiments. A sharp superconducting transition with transition width 0.1 K is seen at 7.5 K in the temperature dependent magnetization measurement. The ZF $\mu$SR spectra show no additional contribution in the relaxation rate below $T_c$, excluding the possibility of time-reversal symmetry breaking in the superconducting state. The Gaussian muon depolarization rate $\sigma_{sc} (T)$ obtained after the analysis of TF spectra could be reconstructed well with a single gap $s$-wave model.  The field dependence of muon depolarization rate $\sigma(B)$ is analyzed using modified London model for single gap $s$-wave symmetry. 
\section{Acknowledgments}
The present work has been carried out at the Swiss Muon Source (S$\mu$S), Paul Scherrer Institute (PSI, Switzerland). The research work of RG was supported by the Swiss National Science Foundation
(SNF-Grant No. 200021-175935). Y. P. Qi is supported by the Natural Science Foundation of China (Grant No. U1932217 and 11974246). The research at Tokyo Tech was supported by the MEXT Element Strategy Initiative to form a research center
(Grant No. JPMXP0112101001).


\begin{thebibliography}{38}
\bibitem{Bednorz}
J. G. Bednorz, and K. A. Muller, Z. Phys. B {\bf64}, 189 (1986).
\bibitem{Steglich}
F. Steglich, J. Aarts, C. D. Bredl, W. Lieke, D. Meschede,W. Franz, and H. Schafer, Phys. Rev. Lett. {\bf43},
1892 (1979).
\bibitem{Kamihara}
Y. Kamihara, T. Watanabe, M. Hirano, and H. Hosono, J. Am.
Chem. Soc. {\bf 130}, 3296 (2008).
\bibitem{Mathur}
N. D. Mathur, F. M. Grosche, S. R. Julian, I. R. Walker, D. M. Freye,
R. K. W. Haselwimmer, and G. G. Lonzarich G G, Nature {\bf 394}, 39 (1998).
\bibitem{Chen0}
G. F.  Chen, Z. Li, D. Wu, G. Li, W. Z. Hu, J. Dong, P. Zheng,
J. L. Luo, and N. L. Wang, Phys. Rev. Lett. {\bf100} 247002 (2008).
\bibitem{Morosan}
E. Morosan, H. W. Zandbergen, B. S. Dennis, J. W. G. Bos,
Y. Onose, T. Klimczuk, A. P. Ramirez, N. P. Ong, and R. J. Cava, Nat. Phys. {\bf2}, 544 (2006).
\bibitem{Yuan0}
H. Q. Yuan, F. M. Grosche, M. Deppe, C. Geibel, G. Sparn, and
F. Steglich, Science {\bf302}, 2104 (2003).
\bibitem{Anderson}
P. W. Anderson, Phys. Rev. B {\bf 30}, 4000 (1984).
\bibitem{Bonalde}
I. Bonalde, W. B.-Escamilla, and E. Bauer, Phys. Rev. Lett. {\bf 94}, 207002 (2005).
\bibitem{Mukuda}
H. Mukuda, T. Fujii, T. Ohara, A. Harada, M. Yashima, Y. Kitaoka, Y. Okuda, R. Settai, and Y. Onuki, Phys. Rev. Lett. {\bf 100}, 107003 (2008).
\bibitem{Bauer1}
E. Bauer, G. Rogl, X.-Q. Chen, R. T. Khan, H. Michor, G. Hilscher, E. Royanian, K. Kumagai, D. Z. Li, Y. Y. Li, R. Podloucky, and P. Rogl, Phys. Rev. B {\bf 82}, 064511 (2010).
\bibitem{Bauer2}
E. Bauer, C. Sekine, U. Sai, P. Rogl, P. K. Biswas, and A. Amato, Phys. Rev. B {\bf 90}, 054522 (2014).
\bibitem{Yuan}
H. Q. Yuan, D. F. Agterberg, N. Hayashi, P. Badica, D. Vandervelde, K. Togano, M. Sigrist, and M. B. Salamon, Phys. Rev. Lett {\bf 97}, 017006 (2006).
\bibitem{Nishiyama}
M. Nishiyama, Y. Inada, and G.-Q. Zheng, Phys. Rev. Lett. {\bf 98}, 047002 (2007).
\bibitem{Chen}
J. Chen, L. Jiao, J. L. Zhang, Y. Chen, L. Yang, M. Nicklas, F. Steglich, and H. Q. Yuan, New. J. Phys. {\bf 15}, 053005 (2013).
\bibitem{Kuroiwa}
S. Kuroiwa, Y. Saura, J. Akimitsu, M. Hiraishi, M. Miyazaki, K. H. Satoh, S. Takeshita, and R. Kadono, Phys. Rev. Lett. {\bf 100}, 097002 (2008).
\bibitem{Das}
D. Das,1, Ritu Gupta, A. Bhattacharyya, P. K. Biswas, D. T. Adroja, and Z. Hossain, Phys. Rev. B {\bf 97}, 184509 (2018).
\bibitem{Rustem1}
R. Khasanov, I. L. Landau, C. Baines, F. La Mattina, A. Maisuradze, K. Togano, and H. Keller, Phys. Rev. B {\bf 73}, 214528 (2006).
\bibitem{Badica}
P. Badica, S. S.-S Jr, A. D. Alvarenga, and G. Jakob, Supercond. Sci. Technol. {\bf 23}, 105018 (2010).
\bibitem{Barker}
J. A. T. Barker, , D. Singh, A. Thamizhavel, A. D. Hillier, M. R. Lees, G. Balakrishnan, D. McK. Paul, and R. P. Singh, Phys. Rev. Lett. {\bf 115}, 267001 (2015).
\bibitem{Singh}
R. P. Singh, A. D. Hillier, B. Mazidian, J. Quintanilla, J. F. Annett, D. McK. Paul, G. Balakrishnan, and M. R. Lees, Phys. Rev. Lett. {\bf 112}, 107002 (2014).
\bibitem{Singh1}
D. Singh, J. A. T. Barker, A. Thamizhavel, D. McK. Paul, A. D. Hillier, and R. P. Singh, Phys. Rev. B {\bf 96}, 180501 (2017).
\bibitem{Karki}
A. B. Karki, Y. M. Xiong, I. Vekhter, D. Browne, P. W. Adams, D. P. Young, K. R. Thomas, J. Y. Chan, H. Kim, and R. Prozorov, Phys. Rev. B {\bf 82}, 064512 (2010).
\bibitem{Bonalde1}
I. Bonalde, H. Kim, R. Prozorov, C. Rojas, P. Rogl, and E. Bauer, Phys. Rev. B {\bf 84}, 134506 (2011).
\bibitem{Hafliger}
P. S. H$\ddot{a}$fliger, R. Kahsanov, R. Lortz, A. Petrovi$\acute{c}$, K. Togano, C. Baines, B. Graneli, H. Kellar, J. Supercond. Nov. Magn. {\bf 22}, 337-342 (2009).
\bibitem{Ying}
T. P. Ying, Y. P. Qi, and H. Hosono, Phys. Rev. B {\bf 100}, 094522 (2019).
\bibitem{Suter_MuSRFit_2012}
A. Suter, and B. M. Wojek, Phys. Procedia {\bf 30}, 69 (2012).
\bibitem{Tinkham}
M. Tinkham, Introduction to Superconductivity (Krieger, Malabar, FL, 1975).
\bibitem{Rustem_AuBe}
R. Khasanov, R. Gupta, D. Das, A. Leithe-Jasper, and E. Svanidze, Phys. Rev. B {\bf 102}, 014514 (2020).
\bibitem{Hirschfeld}
P.J. Hirschfeld, W.O. Putikka, and D.J. Scalapino, Phys. Rev. B {\bf 50}, 10250 (1994).
\bibitem{Serventi}
S. Serventi, G. Allodi, R. De Renzi, G. Guidi, L. Roman`o, P. Manfrinetti, A. Palenzona, Ch. Niedermayer,
A. Amato, and Ch. Baines, Phys. Rev. Lett. 93, 217003
(2004).

\bibitem{Kubo}
R. Kubo, Hyperfine Interact. {\bf8}, 731 (1981).
\bibitem{Bauer3}
E. Bauer, R. T. Khan, H. Michor, E. Royanian, A. Grytsiv, N. M.-Koblyuk, P. Rogl, D. Reith, R. Podloucky, E.-W. Scheidt, W. Wolf, M. Marsman, Phys. Rev. B {\bf 80}, 064504 (2009).
\bibitem{Smidman}
M. Smidman, A. D. Hillier, D. T. Adroja, M. R. Lees, V. K. Anand, R. P. Singh, R. I. Smith, D. M. Paul, and G. Balakrishnan, Phys. Rev. B {\bf 89}, 094509 (2014).

\end{thebibliography}
\end{document}